\documentclass[journal]{IEEEtran}
\usepackage{booktabs} 

\usepackage{times}
\usepackage[scaled=0.92]{helvet}
\usepackage{balance}
\usepackage{microtype}
\usepackage{epsfig}
\usepackage{caption}
\usepackage{subcaption}
\usepackage[T1]{fontenc}
\usepackage{hyperref}
\usepackage{url}
\usepackage{color}
\usepackage{amsmath}

\ifCLASSINFOpdf
\else
\fi
\DeclareCaptionLabelFormat{andtable}{#1~#2  \&  \tablename~\thetable}

\newcommand{\gf}{GreyFiber}
\newcommand{\edit}[1]{\textcolor{black}{#1}}
\newcommand{\editJocn}[1]{\textcolor{black}{#1}}

\newcommand{\squishlist}{
 \begin{list}{${\bullet}$}
  { \setlength{\itemsep}{0pt}
     \setlength{\parsep}{1pt}
     \setlength{\topsep}{1pt}
     \setlength{\partopsep}{0pt}
     \setlength{\leftmargin}{1em}
     \setlength{\labelwidth}{0.5em}
     \setlength{\labelsep}{0.5em} } }

\newcommand{\squishend}{
  \end{list}  }

\begin{document}
%
\title{\gf: A System for Providing Flexible Access to Wide-Area Connectivity}
%
%
%

\author{\IEEEauthorblockN{Ramakrishnan Durairajan\IEEEauthorrefmark{2}, Paul Barford\IEEEauthorrefmark{1}\IEEEauthorrefmark{3}, Joel Sommers\IEEEauthorrefmark{4}, and Walter Willinger\IEEEauthorrefmark{5}}\\
\IEEEauthorrefmark{2}University of Oregon,
\IEEEauthorrefmark{1}University of Wisconsin-Madison,
\IEEEauthorrefmark{3}comScore, Inc,
\IEEEauthorrefmark{4}Colgate University,
\IEEEauthorrefmark{5}NIKSUN, Inc.}

\markboth{}%
{Durairajan \MakeLowercase{\textit{et al.}}: \gf: A System for Providing Flexible Access to Wide-Area Connectivity}
%



\maketitle

\begin{abstract} 
Access to fiber-optic connectivity in the Internet is traditionally offered either via lit circuits or dark fiber. Economic (capex vs. opex) and operational considerations (latency, capacity) dictate the choice between these two offerings, but neither may effectively address the specific needs of \editJocn{modern-day enterprises or service providers} over a range of use scenarios. In this paper, we describe a new approach for fiber-optic connectivity in the Internet that we call \gf. The core idea of \gf~is to offer flexible access to fiber-optic paths between end points \editJocn{({\em e.g.}, datacenters or colocation facilities)} over a range of timescales. We identify and discuss operational issues and systems challenges that need to be addressed to make \gf~a viable and realistic option for offering flexible access to infrastructure (similar to cloud computing). We investigate the efficacy of \gf~with a prototype implementation deployed in the GENI and CloudLab testbeds. Our scaling experiments show that 50 circuits can be provisioned within a minute. We also show that backup paths can be provisioned 28 times faster than an OSPF-based solution during failure/maintenance events. Our experiments also examine \gf~overhead demands and show that the time spent in circuit creation is dependent on the network infrastructure, indicating avenues for future improvements.
\end{abstract}


\IEEEpeerreviewmaketitle

\vspace{-0.25cm}
\section{Introduction} \label{sec:introduction} 

\IEEEPARstart{T}{he} premise of cloud computing is that instead of building and maintaining their own in-house computing and storage infrastructures, users ({\em e.g.}, companies, organizations, individuals) can consider and consume compute resources as a utility. ``The network is the computer'' is a much-used phrase that was coined to succinctly describe this utilitarian approach that has been the driving force for much of the ongoing ``cloudification'' of today's Internet. For users, the benefits of relying on the network to perform tasks that traditionally ran in local compute environments are all too obvious. 
\editJocn{Specifically, the ability to spin up compute resources on demand for virtually any workload offers enormous flexibility and reduced operational complexity for users. Indeed, the commonly-adopted pay-per-use billing model for cloud computing has proven to be a key economic incentive driving the recent emergence of cloud-related ecosystems ({\em i.e.}, cloud providers and services), technologies ({\em e.g.}, Software-Defined everything Infrastructure (SDxI)~\cite{sdxCentral}) and paradigms ({\em e.g.}, Cloud 3.0 and BigCompute~\cite{cloud3dot0}). }

However, these benefits also have a direct impact on the type of traffic that is generated in a \editJocn{Cloud 3.0-centric Internet} and on how that traffic is routed over the existing physical Internet infrastructure (see for example~\cite{klinkowski2013advantages,cflam2010}). Consider for example the simple case of different users spinning up virtual machines (VMs) for running big data analytics applications that require the transfer of large datasets from a geographically-dispersed set of datacenters (DCs), possibly with additional performance- or security-related requirements ({\em e.g.}, low-delay, resilience to outages, avoiding certain networks or regions). Such transfers can potentially consume significant portions of the available bandwidth along their routes, but the onus is squarely on the user's cloud provider or on that cloud provider's transit provider to ensure that the user's application gets the necessary data as required. Traditionally, traffic engineering and routing have been used to address such issues ({\em e.g.}, see~\cite{liu2015traffic,zhang2015guaranteeing,kandula2014calendaring,danna2012practical,laoutaris2011inter,kandula2005walking,applegate2003making,fortz2000internet}), but what if the nature of the generated traffic is such that it periodically exceeds the available capacity on the primary and backup paths and no alternative paths are available? 

There have been recent efforts to study the problem of dealing with highly variable and unpredictable \editJocn{workloads in inter-DC WANs ({\em e.g.},~\cite{jain2013b4, hong2013achieving, jin2016, singh2017run, channegowda2013software, giorgetti2015dynamic, kilper2017optical, xiong2018sdn})}. In particular, B4~\cite{jain2013b4} and SWAN~\cite{hong2013achieving} leverage SDN technology and rely on a wide area network view to dynamically change routing and rate allocations to ensure high network utilization while meeting the deadlines of the data transfers. However, by assuming a fixed network- or router-level WAN topology, these efforts ignore the opportunities that arise from reconfiguring equipment in the underlying optical or physical layer to dynamically change the router topology. Such a joint (and central) control of both the physical and network layer has recently been considered in~\cite{jin2016} where the authors describe Owan, a new SDN-based system for orchestrating bulk transfers that computes and implements the optical circuit configuration ({\em i.e.}, the optical circuits that implement the network-layer topology) and the routing configuration ({\em i.e.}, the paths and rate allocation for each transfer) to ensure high network utilization and optimize bulk transfers.

In this paper, we move beyond~\cite{jin2016} and borrow a page from cloud computing. In particular, we describe the design and implementation of {\it \gf}, a new platform for establishing fiber-optic connectivity in the Internet. Similar to how the cloud enables arbitrary users to spin up VMs as needed, \gf~makes it possible for infrastructure providers to spin up optical circuits on demand to handle the highly variable and unpredictable workloads that a cloud-centric Internet entails. In a sense, \gf~is to the wide-area Internet as 3D beamforming is to DCs~\cite{Zhou2012}. While the technologies, economics, and operations underlying these two approaches differ drastically, their objectives are the same. That is, to alleviate traffic hotspots as they occur as the result of highly unpredictable traffic, the original (fixed) means of data communication is complemented by unused communication channels that are made available as needed---idle optical circuits in the case of \gf~in the wide-area Internet, and idle wireless links in the 60 GHz band for 3D beamforming in DCs. In fact, where available, \gf~could include the sort of microwave communication that is used for high-frequency trading applications between New York and Chicago~\cite{ref-hft1,ref-hft2} (see also~\cite{speed-of-light-internet})\footnote{In this paper, our focus is on utilizing unused optical circuits.}.

The main idea for \gf~is to provide a means to offer easy and cost-effective access to unused fiber-optic paths between participating endpoints \editJocn{({\em e.g.}, datacenters and/or colocation facilities)} on demand, for arbitrary durations, and possibly with industry-specific performance guarantees ({\em e.g.}, ultra-low delay for high-frequency trading applications or gaming services; fully diverse physical paths for mission-critical business applications). In this sense, \gf~can be thought of as offering {\em wide area connectivity as a service} \editJocn{and as a {\em realization of Bandwidth on Demand} (BWoD~\cite{bwod})---one of the key cornerstones fueling software-centric innovations in Cloud 3.0}. However, \gf~differs from standard cloud computing services ({\em e.g.,} SaaS, PaaS and IaaS) in that it is fundamentally concerned with connectivity, not computation. \edit{In the rest of the paper, we use the following terminology. The unit of connectivity in \gf~is a {\em link} which refers to a single strand of fiber. A link may contain one or more {\em circuits}, which are defined as logical connections across endpoints with unique wavelengths and which are configurable sub-units in \gf. Multiple links are bundled in a {\em path} (also known as a {\em conduit}) and each path/conduit is physically installed between endpoints at distinct geographic locations.}

The design of \gf~requires the careful integration of three critical components. First, to ensure that \gf~is an economically viable option, we monetize the current over-supply of buried fiber in existing conduits in today's physical Internet infrastructure~\cite{fiberGlut1,fiberGlut2}\footnote{Our focus here is strictly US-centered.} by proposing an auction-based Fiber Exchange that attracts potential buyers and sellers of \gf. Second, we leverage the fact that fiber-optic technology has advanced to the point where today's fiber-optic gear allows fast remote reconfigurations. For example, provisioning of an idle circuit can be done on the order of milliseconds to seconds~\cite{chiu2012architectures,oclara,infineraDatasheet,infinerasdn,ovum} which suggests that spinning up an optical circuit between two participating endpoints can be achieved at time scales that are commensurate with those required for launching a cloud service. Finally, the operation of our \gf~platform is inspired by prior work~\cite{jin2016} and relies on a central controller that allows for direct and end-to-end control of all \gf-affected devices and simplifies overall network management.

To demonstrate the feasibility of our approach and examine its efficacy, we first describe an implementation of our \gf~design and deploy it in the GENI testbed. This prototype system addresses the technical challenges associated with circuit provisioning and enables performance evaluation over a range of use scenarios. In particular, we show that as many as 50 paths can be provisioned between endpoints in less than a minute, which demonstrates the rapid provisioning capabilities of \gf. To enable higher infrastructure resilience during network outages and/or planned maintenance events, we also show how \gf~can be used to create an effective backup solution. Specifically, \gf~can reactively detect path failures and provision a new path within 1.25s, which outperforms the traditional OSPF-based backup solution by 28x. This agility of \gf~benefits many applications by allowing them to be oblivious to underlying network failures. Finally, we dynamically provision paths between endpoints to create on-demand high-capacity connectivity and demonstrate the resulting performance benefits of GreyFiber. In addition, to show that \gf~is adaptable to different networking substrates, we also report on an experiment that leverages CloudLab~\cite{cloudlab} and demonstrates \gf's ability to scale to high-bandwidth links.

We quantify the overhead of our system versus the underlying infrastructure and highlight the critical path performance of \gf. By examining the log files produced during circuit provisioning, our analysis shows that \gf~has minimal system overheads. We find that the latency overhead for on-demand path provisioning is completely dependent on the underlying network substrate ({\em e.g.}, hardware), which highlights avenues for improvement and expansion of the range of use scenarios of \gf~in the future.

\vspace{-0.35cm}
\section{The Case for \gf} \label{sec:motivation} 

Over the past several years, there have been significant changes among network service and infrastructure providers that motivate the timeliness of {\em wide-area connectivity as a service} embodied in \gf.

\textbf{Consolidation of dark fiber providers.} There has been a trend toward consolidation among dark fiber providers.  Examples include CenturyLink's acquisition of Qwest in '11 (resulting in a combined 190k mile fiber network~\cite{centurylink}), Zayo's acquisition of Abovenet in '12 (resulting in a combined 6.7M fiber mile network connecting some 800 datacenters~\cite{zayo}),
Lighttower merging with Fibertech in '15~\cite{acq2015},  CenturyLink's acquisition of Level 3 in '16 ~\cite{acq2016CL}, and Verizon's recent announcement to acquire XO communications' fiber-optic network business~\cite{acq2016}.  A clear consequence of these mergers is that \textit{there are fewer fiber-optic network providers, but the remaining ones have larger fiber footprints}.


\textbf{Evolution in the datacenter market.} There has been consolidation as well as expansion within the datacenter market. Among the tier-1 datacenter providers ({\em i.e.}, serving major metro areas and large cities), examples of consolidation include Equinix' acquisition of Telecity Group (EU/UK)~\cite{equinix1} and Bit-Isle (JP)~\cite{equinix2} in '15,
AT\&T announcement to sell datacenter assets~\cite{att} in '15 and Windstream's announcement to sell its datacenter business to TierPoint~\cite{windstreamdc} in '15. At the same time, the growing demand for cloud services has put pressure on the largest cloud providers to have presence in more locations and also closer to their customers, which has led to the emergence of an increasing number of new 2nd-tier datacenter providers ({\em e.g.,} EdgeConneX~\cite{EdgeCoX}) that are focused on medium-sized markets such as Portland, OR and Pittsburgh, PA. The combined effects of this cloud-driven, broader user-base and higher volatility of workloads could be mollified via \gf. These trends indicate an \textit{expanding geographic distribution of datacenter capacity that could benefit from \gf~connectivity}.

\textbf{Dark fiber providers acquiring datacenters.} There are recent examples of dark fiber providers acquiring datacenters, which presents an opportunity for one provider to supply high-bandwidth connectivity between datacenter co-location endpoints to customers who need it.  One example of a provider with this nascent capability is Lightower~\cite{lightower}, which acquired ColocationZone in '15~\cite{lightowerdc2} and Datacenter101~\cite{lightowerdc1} in '16. Similarly, Allied Fiber\footnote{Allied Fiber is now defunct primarily because they were not able to build an adequate customer base quickly enough~\cite{articleaboutalliedfailing}.} aimed to be a network-neutral and dark fiber ``superstructure'' with  a footprint across the United States and offered traditional 20-year and non-traditional 12, 24, and 36-month Indefeasible Rights of Use (IRU) options~\cite{alliedfibertiers}. {\em These developments indicate that there exist business opportunities for companies that offer integrated (network-neutral) colocation/dark fiber services and that could benefit from available \gf~connectivity to boost their existing but maybe constrained dark fiber infrastructure.}

\textbf{Implementation challenges.} Our framework for \gf~includes three high-level aspects:  a fiber exchange, a circuit provisioning system and a central controller.  Each component has its own technical challenges to enable scalable use across diverse physical infrastructures.  In most respects, the fiber exchange has the same requirements as other auction-based systems ({\em e.g.,} Amazon EC2 spot pricing system~\cite{EC2}), and indeed those provide a blueprint for our \gf~prototype described in \S\ref{sec:methods}.  Next, driven by demands in datacenters, new optical switching equipment is being designed to speed and simplify configuration and management of optical connections~\cite{ovum}.  For example, Infinera's Open Transport Switch~\cite{infinerasdn} is a software layer that runs on top of of their optical cross connect hardware to enable fiber-optic wavelengths to be put into service on demand. We believe that this trend in switch technology, which is a key enabler for \gf, will continue in the future.
Finally, the global controller must coordinate between user requirements and the underlying physical infrastructure to ensure that service commitments are satisfied.  These requirements are akin to SDN controllers, which serve as a model for our \gf~prototype (\S\ref{sec:methods}).

\textbf{Incentives for \gf.} While corporate and technical trends indicate the opportunity for \gf, practical incentives motivate broader deployment and use.  We consider the incentives for \gf~versus IP transit ({\em i.e.}, lit fiber) and dark fiber, which are the standard fiber options in the Internet today. In particular, we compare and contrast the three market options using five different metrics: economic incentives; potential market size; control over routing; physical route diversity and control over performance.

Table~\ref{tab:incentiveComparison} shows a relative comparison between the \gf~and other fiber markets.  Based on the IP-transit and dark fiber price sheets compiled from three different US service providers, we posit that dark fiber has the lowest economic incentive if one considers a broad set of customers.  First, there is the required 20+ year commitment for an IRU, which locks in capital expenditures (CAPEX) and operational expenditures (OPEX) over that duration.  The standard pricing model for dark fiber includes an upfront payment for the IRU along with substantial CAPEX to light fiber. Reoccurring costs include CAPEX at $\sim$\$1000 to 3000 per mile per year and OPEX at  $\sim$\$250 per mile per year. These costs and the duration of the commitment tend to reduce market size.  Benefits of dark fiber include control over routing, physical route diversity and low latency due to direct interconnection to peers at the colocation facilities.

\vspace{-0.4cm}
\begin{table}[htb!]
\centering
\caption{Incentives of \gf-based fiber market in comparison with the IP transit and dark fiber options (L = low, M = medium, H = high).}
\vspace{-0.3cm}
\label{tab:incentiveComparison}
\scalebox{0.8}{
\begin{tabular}{l|ccc}
                         			& Dark Fiber& IP Transit  & \gf 		   \\ \hline
Economic Incentives		& L & M & H        \\
Potential market size		& L & M & H       \\
Control over routing		& H & L & M        \\
Physical route diversity  	& M & L & H        \\
Control over performance 	& M & M & M
\end{tabular}
}
\end{table}
\vspace{-0.4cm}

Fiber pricing in the IP transit market is $\sim$\$500--600 per Gbps per month.  Benefits include medium-term commitments (3--5 years) for fully managed services, no OPEX or CAPEX. The one-stop shopping, fully managed service aspect of IP transit leads to a medium sized market. The drawbacks include ({\em 1}) no physical route diversity (unless explicitly specified at additional cost), ({\em 2}) no routing control, and ({\em 3}) latency determined by SLA, which may be insufficient due to indirect routing and lack of direct interconnection at peering points.

In this paper, we assume \gf~will initially be offered in auction-based exchanges as managed layer-3 services.  Thus, the benefits of \gf~include {\em (1)} a flexible pay-as-you-go model and no upfront costs, which opens the fiber market to a potentially large customer base; {\em (2)} the ability to choose diverse routes; and {\em (3)} control over performance ({\em i.e.}, low latency) due to direct interconnection with peers.  As a consequence, the only drawback is that the customers will likely have limited control over routing.

\textbf{Use cases.} We envision three use cases for \gf: {\em (1)} improving network resilience through redundant connections, {\em (2)} providing (ultra) low latency paths, and {\em (3)} providing on-demand high-capacity paths over arbitrary durations.  Internet outages are common and occur due to a variety of reasons including accidents, misconfigurations and censorship ({\em e.g.,} ~\cite{mcgrattan06,zmijewski08,Philip08,dainotti11}).  Outage can be mitigated by temporary paths that reconnect points within a network.  Addition of a long-haul path might also be considered as a preemptive measure in the case of a planned maintenance outage, or knowledge of an impending weather event that may affect the network.  Next, a reduction of milliseconds or even microseconds in latency can yield competitive advantages in the financial sector or in gaming.  The addition of new fiber links through \gf~may be used to provide more direct paths and thereby reduce end-to-end latency.  Finally, the need to transfer (large) data sets across the wide-area Internet or between datacenters is likely to continue to grow. \edit{Improving throughput and scheduling of large inter-datacenter transfers has been the subject of recent research ({\em e.g.}, Netstitcher~\cite{laoutaris2011inter}, Pretium~\cite{jalaparti2016dynamic}, BWoD~\cite{bwod}, and BDT~\cite{wu2015orchestrating}), and could benefit from additional high-capacity paths via \gf. Furthermore, for large ISPs with over-provisioned backbones, the additional capacity needed at a given network location and at a given time may not be instantly available. Similarly, the smaller providers may not have large and/or over-provisioned backbones and may need additional capacity at different places in time. In summary, {\em \gf~is a flexible alternative to standard lit/dark fiber options with both economic and operational advantages.}}

For each of these motivating use-cases, there may be quite different requirements in terms of capacity and the time duration over which the additional capacity is needed.  For example, ({\em i}) short lifetime capacity to address an unexpected outage, ({\em ii}) short lifetime capacity to address unexpected demand, ({\em iii}) short lifetime capacity to enable better performance between two points, ({\em iv}) medium lifetime capacity to service expected demand that has no specific deadlines, ({\em v}) short-to-medium lifetime capacity for transit, backhaul, etc.  We believe that these scenarios create a compelling case for the utility of on-demand connectivity offered by \gf.


\vspace{-0.35cm}
\section{\gf~System Design} \label{sec:methods} 


\subsection{System Requirements}
\vspace{-0.1cm}
\label{req}
A \gf~system must satisfy the following requirements:

{\bf Scalability and extensibility.}  The system must scale to meet the demands of envisioned sellers and buyers.  From sellers' perspectives, this could mean providing access to many thousands of circuits driven by diverse hardware across a broad geographic region. From buyers' perspectives, this means having access to potentially thousands of end-to-end paths that are available in many/most colocation facilities in a broad geographic region.  Further, the fiber exchange must scale to meet diverse demands of buyers in a timely fashion.

{\bf High availability.} Service/content providers typically seek to guarantee five-nines availability to their customers~\cite{Govindan2016} ({\em i.e.}, available 99.999\% of the time). Likewise, the \gf~system must be highly available in order to function as a flexible provider of wide-area connectivity.  Additionally, the resources ({\em i.e.}, endpoints and links) provisioned by the system should also enable/support five-nines availability.  Two positive consequences of such a highly-available system are that failures can be treated as a normal situation to be handled,
and that a high level of service can be guaranteed through service level agreements (SLA), with low risk to the provider.

{\bf Rapid provisioning.}  Hardware resources must be able to be provisioned over short timescales (ideally on the order of millisecond or submillisecond).  This capability enables \gf~paths to be available over very short timescales ({\em e.g.}, in response to workload bursts) and to put paths into service quickly when needed by a customer to recover from an unexpected failure.  Naturally, for ISPs, a fast infrastructure provisioning capability simplifies the process of activating backup resources during network maintenance or outage events.  Rapid provisioning also implies the need for a system that is easy-to-use after it has been initially configured.

{\bf Flexible access.}  Current dark fiber leases (based on 20+ year IRUs) and IP transit commitments (typically 3-5 years) inherently limit access to connectivity.  To overcome this impediment \gf~requires access to infrastructure over a wide range of  timescales (sub-second to years).  This enables many opportunities for buyers and sellers including economic benefits, reselling unused resources, and ease of expansion at diverse geographical regions.

\vspace{-0.38cm}
\subsection{\gf~Overview}
\label{subsec:overview}
\begin{figure*}[!tbp]
  \centering
  \begin{minipage}[b]{0.35\textwidth}
    \includegraphics[width=\textwidth]{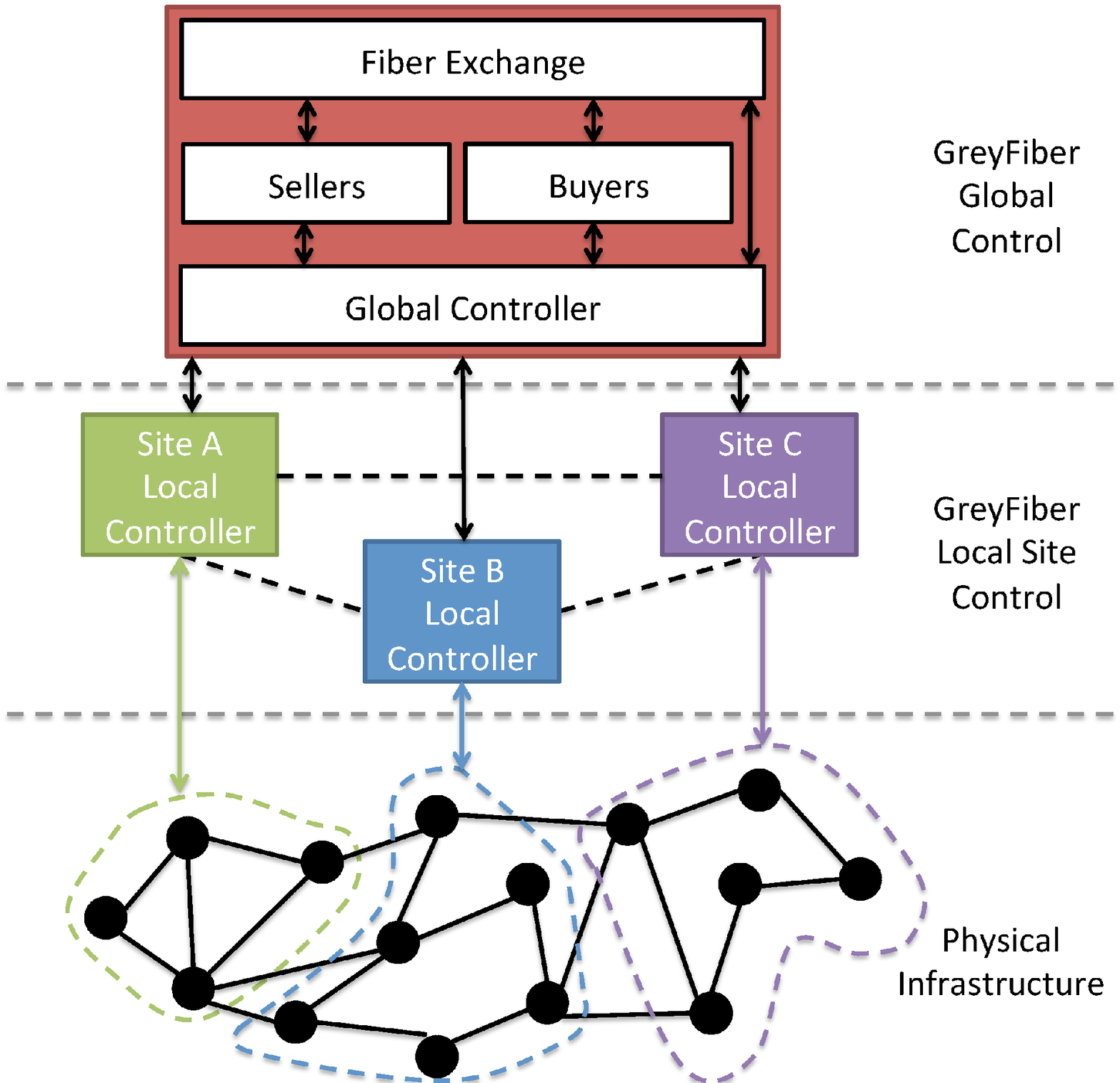}
    \vspace{-0.5cm}
    \caption{\gf~architecture.}
    \vspace{-0.4cm}
    \label{fig:gfArchitecture}
  \end{minipage}
  \hspace{0.5cm}
  \begin{minipage}[b]{0.6\textwidth}
    \includegraphics[width=\textwidth]{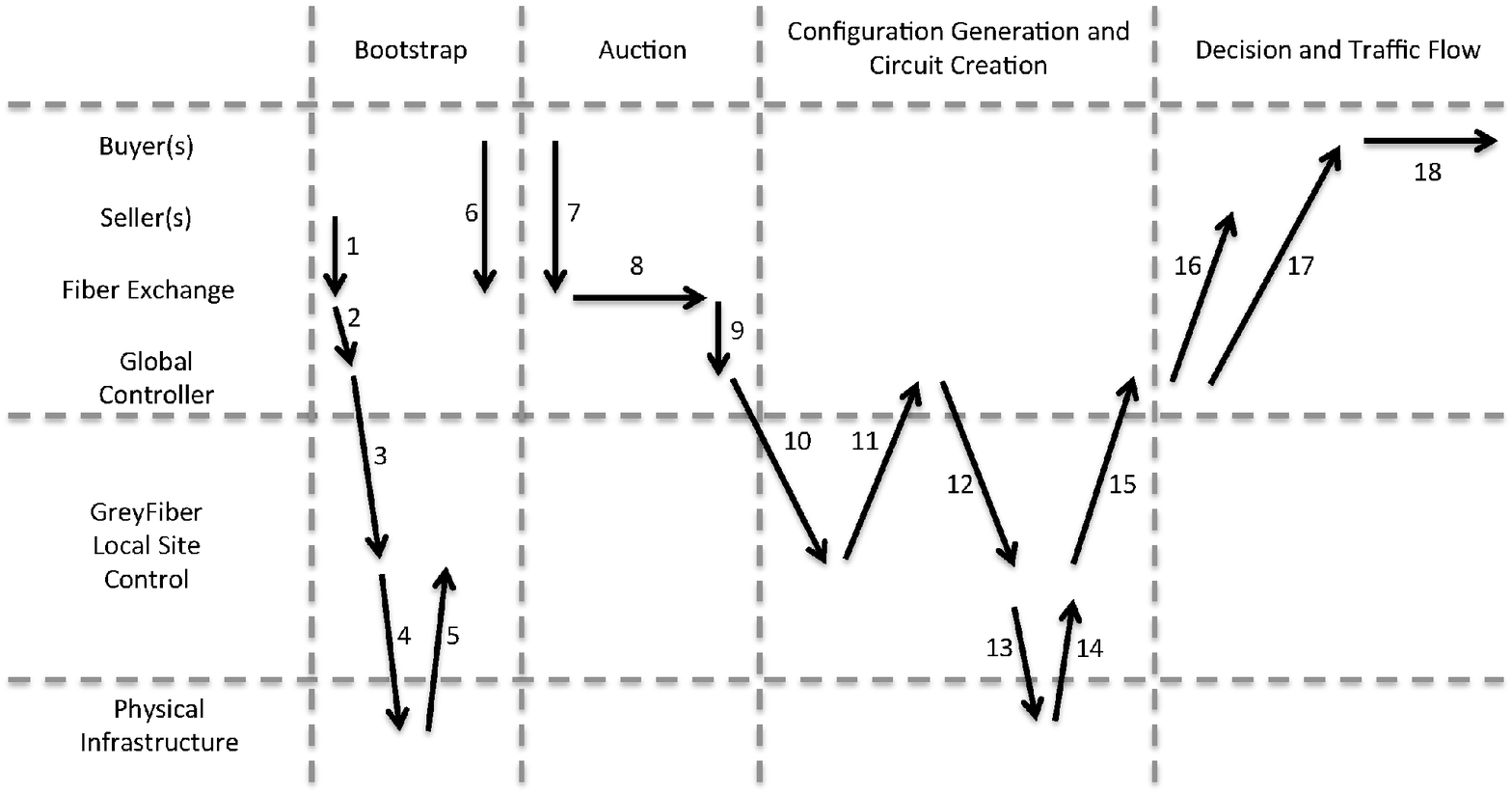}
    \vspace{-0.5cm}
    \caption{\gf~timeline of events to enable end-to-end connectivity.}
    \vspace{-0.4cm}
    \label{fig:gfEvents}
  \end{minipage}
\end{figure*}

\gf~is a three-tiered system whose goal is to provide wide area connectivity as a service over a range of timescales. \gf~consists of three components: {\em (1)} Global Control, {\em (2)} Local Site Control, and {\em (3)} Physical Infrastructure. The overall architecture of \gf~is depicted in Figure~\ref{fig:gfArchitecture}, which is inspired by the hybrid control proposed by~\cite{mukerjee2015practical}. 

{\bf \gf~Global Control.} The highest level of the system is the \gf~Global Control (GGC), which serves as a command center for the entire system by providing a common interface for all the entities involved. To meet scalability, extensibility and availability requirements, the GGC resides either in the cloud or in a datacenter and consists of the following four sub-components/entities:
\squishlist
\item {\em Fiber Exchange.} An auction management system (explained in \S\ref{auctionmodel}) that is similar to an ad auction~\cite{abrams2007ad,adwords}
or cloud resource auction system~\cite{zheng2015bid}. This subcomponent can either be co-located with the Global Controller or can reside in a different location ({\em e.g.}, the cloud).
\item {\em Buyers.} The entities ({\em e.g.}, ISPs, CDNs, enterprise networks, etc.) or the customers of \gf~who specify their connectivity needs---also known as {\em resource requests}---including geography, performance, timescales, deadlines (if any) and bandwidth requirements, along with their bids are called {\em Buyers}. Support for these options allows planning over longer time scales where buyers can manage costs ({\em i.e.}, leasing vs. digging new conduits) or over short time scales when there is a specific need ({\em i.e.}, during specific Internet events like high-traffic streaming events, planned outages due to maintenance, etc.).
\item {\em Sellers.} The entities ({\em e.g.}, service, cable and fiber providers) who own/have the ability to provide a link or set of links to the \gf~ecosystem are called {\em Sellers}. To support evolution in the physical Internet and to enable a \gf-based connectivity service, an entity has to meet the following constraints: {\em (i)} provide access to all (layer 1) hardware such as endpoints and links, {\em (ii)} provide access to the routing substrate in order to direct packet traffic to the lit fiber, and {\em (iii)} support for a wrapper API to get circuit provision/tear down decisions from the auction-based decision process.  We call these three constraints {\em seller requirements}. Similar to any market with competing entities, we hypothesize that different sellers compete based on factors including fiber costs, geographical diversity and robustness of their paths, and simplicity in establishing/tearing down connectivity.
\item {\em Global Control.} A centralized controller (similar to an SDN controller) that has a global view of all site controllers, also known as GLSCs (explained below), at different geographic locations. Various applications including traffic engineering, time-based circuit provisioning, network management and backup restoration are implemented within this entity.
\squishend

In \gf, access to connectivity is based on winning auctions for available resources ({\em e.g.}, either via generalized second price auction (GSP)~\cite{edelman2007internet} or Vickrey-Clarke-Groves (VCG)~\cite{Vickrey1961,Clarke1971,Theodore1973} auctions). Winning a bid results in a configuration that is pushed instantaneously across the sites for a specified customer. Circuit creation is similar to the flow installation using a circuit pusher~\cite{circuitpusher} application in FloodLight. A wide variety of time-based circuit provisioning capabilities (\S\ref{operationaltaxonomy}) are also supported in \gf.


{\bf \gf~Local Site Control.} Below the GGC is the \gf~Local Site Control (GLSC) which mimics minimal functionalities from the GGC in a local context ({\em e.g.}, local decisions on failures, provisioning next available resource in case of failure, etc.) and provides local control for individual sites at marked geographic locations. With the rise of Internet Exchange Points (IXP), researchers have observed a ``flattening'' of the peering structure in the Internet~\cite{Gill08, Dhamdhere10, Labovitz10}, affecting the structure of end-to-end paths; these facilities are natural locations for GLSCs.  Accordingly, we assume that GLSCs are available in every colocation facility.  A GLSC has the following capabilities:

\squishlist
\item {\em register} with the Fiber Exchange where the registration includes information about the set of links, capacity required, geographic reach and the potential buyers that are directly connected to a particular GLSC;
\item {\em configure} links, that is, when a buyer wins an auction, connectivity is established for the specified period of time over the specified link(s) and then to tear down these connections when the time expires;
\item {\em report status} information to the exchange since the link may not always be available or buyers might be interested in real time status, especially on links that are used by multiple buyers;
\item {\em control} a set of physical infrastructure (explained below) during connection setup and tear down; and
\item {\em monitor} links connected to them and maintain different performance indicators like packet loss, latency, and connection stability.
\squishend

In the future, we envision replacing these GLSC units with either SDN-enabled IXPs or simply Software-Defined Exchanges (SDX)~\cite{gupta2014}, where inside an SDX, the route servers are local SDN controllers and SDN-enabled switches where multiple ASes participate, connect and exchange traffic.

{\bf Physical Infrastructure.} The final layer in the \gf~ecosystem is the Physical Internet Infrastructure, which is composed of traditional nodes (fiber connection points) and links (fiber strands)~\cite{durairajan2013, durairajan2015}. The physical Internet layer encompasses both long-haul and metro fibers, which provide intra- and inter-GLSC connectivity.  Although we conceive of this layer as {\em physical} infrastructure, any network substrate for which the required GLSC functions can be implemented can fulfill this role, {\em e.g.}, overlay or virtual network topologies created using
Mininet~\cite{mininet} or GENI~\cite{Berman20145}.

\edit{It is important to note that there are many {\em technical} and {\em engineering} challenges that must be overcome at the physical layer to realize rapid connection setup/teardown.  Technical issues include signaling across various endpoints, hardware limits such as transmission power, and fiber-specific challenges such as attenuation and chromatic and polarization mode dispersion~\cite{Slide39}.
In this paper, we assume that these factors are already addressed and that the Sellers expose the configurable wavelengths of fiber strands (as part of Seller requirements) to GGC to ensure that the wavelengths are unique for each created circuit.
Moreover, since the signal-to-noise ratio of other wavelengths is affected when a new wavelength is added dynamically, the optical power needs adjustment every time a new circuit is added.  We plan to consider such power adjustments in future work. In addition, our future efforts will investigate the efficacy of CDC ROADM-based wavelength reconfigurability (\cite[slide 39]{Slide39}) in \gf.}

\edit{Some of the engineering challenges include determining locations for infrastructure build outs, deploying endpoint-specific capabilities ({\em e.g.}, amplifier, multiplexer, signal regeneration equipment, etc.), patching endpoints to fiber strands, and electricity needed to power the deployments. Since the speed at which the \gf~system can put new paths into service is dependent on many factors, including the engineering challenges mentioned above, our requirement is that they not add any significant overhead to the provisioning times of the underlying paths and/or links under its control. Furthermore, we assume that these factors are taken care of at the Sellers' end before using \gf. That is, {\em the fiber path is already lit between endpoints and every seller controls, manages and maintains their own portion of the physical infrastructure.}}

\vspace{-0.3cm}
\subsection{Supported Circuit Provision Scenarios}
\label{operationaltaxonomy}
To overcome the inflexibilities in standard infrastructure leasing (\S\ref{req}) and to address the need for quick, dynamic and on-demand network parallelization and/or circuit provisioning, the GGC in \gf~supports a wide taxonomy of time-based provisioning scenarios. At the highest level, the provisioning module that implements the time-based provisioning logic classifies the {\em resource requests} from buyers into either a {\em realtime} or a {\em non-realtime} request. Once the immediateness of a given request is identified by the provisioning logic, it is further sub-classified based on {\em (i)} timescales during which the path is needed, {\em (ii)} backup requirements, and {\em (iii)} scalability/performance constraints.

\gf~supports a variety of circuit provisioning scenarios at varying timescales including {\em small} (from seconds to minutes), {\em medium} (hours), {\em large} (from days to months) and {\em extra-large} (years similar to a standard fiber lease or IRU). In addition, circuits could be dynamically provisioned to serve as  backups during (or quickly after) either an outage event or a scheduled maintenance operation. Furthermore, in order to meet performance constraints in the SLA at peak times, links could be elastically spun up and/or down using \gf.

\vspace{-0.2cm}
\subsection{Auction Model}
\label{auctionmodel}

To enable flexibility in infrastructure pricing, the GGC---in particular, the Fiber Exchange subcomponent---uses an auction model to lease seller's infrastructure to interested buyers/customers.  Making~\gf~resources available via auction recognizes that the value in wide area connectivity as a service is in the excess capacity available over a variety of time scales (similar to the motivation for spot markets in cloud infrastructures). Should customers wish a longer term IRU, traditional dark fiber and IRU-based leasing model are assumed to be available.

Fiber Exchange offers auctions from a list of $k$ links\footnote{In this work, our key focus is to enable leasing the fiber/link resources. However, there is nothing limiting in \gf~to support leasing of other types of resources ({\em e.g.}, routers).}. Specifically, there is list $L$ of $k$ links, where:
\begin{align*}
L = \{l_{1}, l_{2}, l_{3}, \dots, l_{k}\}
\end{align*}
\vspace{-0.4cm}

There are $N$ ($>$ $k$) customers, each of whom submits one bid per link, a non-negative value $b_i$, independently and simultaneously with other bidders. Note that a customer can bid for multiple links ({\em e.g.}, $l_1$, $l_2$ and $l_3$) separately and a path ($p$) can be a composition of either multiple links (say $l_1$-$l_2$-$l_3$) that is laid sequentially in different conduits or three strands of fiber laid in parallel within the same conduit. In what follows, we explain the generalized second price (GSP) auction~\cite{edelman2007internet}, which is the default resource auction model in \gf.

The auction format is GSP with perfect information, and the selection rule is such that the highest $k$ bidders are ranked by their bid values. The payment that the winner makes is the second-highest bid among those submitted by the players who do not win for a particular link. In such a setting, the payoff function, which also denotes the preference of customer $i$ for a link $l_i$, is given by:
 \begin{displaymath}
   u_i = \left\{
     \begin{array}{ll}
       v_i - \hat{b} & \text{if} ~ b_i \geq \hat{b} ~ \text{and} ~ v_i > v_j  ~ \text{if} ~ b_j = \hat{b} \\
       0 		& \text{if} ~ b_i < \hat{b}
     \end{array}
   \right.
\end{displaymath}
subject to the following (seller) constraint,
 \begin{displaymath}
	v_i \leq b_{i} \quad \forall i={l_1, \dots, l_k} \\
\end{displaymath}

where, each bidder/customer submits a (sealed) bid $b_i$, and $\hat{b}$ is the highest bid submitted by a customer other than $i$. $v_i$ is the value that seller attaches to every link $l_i$ to maintain revenue. In short, if the customer obtains a link, they receive a payoff $v_i$ -- $b_i$. Otherwise, their payoff is zero. Furthermore, the benefits of GSP including enabling a more user-friendly market that is less prone to gaming by other bidders is shown by Edelman {\em et al.}~\cite{edelman2007internet}.

Note that our auction mechanism does not preclude a traditional lease, since a contract could be offered on an exchange with the reserve price set at the standard lease rate.  Therefore, \gf~is backwards compatible. It is possible for a new entrant to use \gf~with short-term leasing option while others use a legacy model with long-term IRU-based leasing. Furthermore, while the idea of applying auction-based methods for leasing a service provider's infrastructure in \gf~is new, the auction mechanisms are well known\footnote{Other forms of auction mechanisms such as GSP with reserve pricing could also be used.}.


\subsection{End-to-end Events in \gf}
{\bf Assumptions.} To establish an end-to-end circuit between endpoints (A and B) of a customer, we assume that the customer has one or more of the following options between their endpoints and a colocation facility that is \gf-enabled: {\em (1)} metro-fiber or broadband or wireless connectivity (and access) in the last mile~({\em e.g.}, Verizon's Interconnection services~\cite{verizonICS}), or {\em (2)} a dedicated private connection ({\em e.g.}, Microsoft's ExpressRoute~\cite{expressRoute}), or {\em (3)} Fibre to the Premises (FTTP) on Demand~\cite{fttpOD}.
Furthermore, we assume that the connectivity between customer endpoints and GLSC units are already lit and tested.

Below, we describe the events that take place to establish end-to-end connectivity,
as shown in Figure~\ref{fig:gfEvents}:
\squishlist
\item Every seller registers with the \gf~system with information that includes the geography of nodes and links\footnote{\edit{Service providers are aware of geographic locations at which other ISPs peer, along with node/fiber footprint. This information is revealed either through documents and filings~\cite{durairajan2015} or through voluntarily data given by the providers~\cite{PeeringDB1}}.}, peering and link properties ({\em e.g.}, capacity, performance indicators). This information is communicated to the Fiber Exchange and is also advertised to a list of buyers in the ecosystem (step 1). Every buyer must also register with the \gf~system prior to entering bids (step 6).
\item Once the registration is complete, the GGC forwards the information to the appropriate GLSCs (steps 2 and 3), which monitor the requested set of links for various performance indicators including latency perceived, loss and link utilization (steps 4 and 5).
\item If there is a demand from a buyer in the form of resource requests, their bids along with other relevant information are accepted (step 7) and the Fiber Exchange runs an auction to determine the winner (step 8).
\item \edit{Once a winner for a link is determined, the Fiber Exchange communicates the winner information and their corresponding requirements to the Global Controller (step 9).}
\item \edit{At the global controller, creation of circuit(s) (in a link) between buyer endpoints occurs in two stages. First, the physical topology graph $G$ is queried (step 10). $G$ is composed of fiber strands from multiple sellers. Each edge in the graph is annotated with maximum and available bandwidth, and total number of fiber strands; if the requested bandwidth and number of fiber strands is admissible, the resource request proceeds to the second stage (step 11), otherwise it is aborted.}
\item \edit{The establishment of an end-to-end circuit happens in the second stage and is composed of multiple events (steps 12 to 15). The logical end-to-end circuit
is stitched from individual
links in $G$. Buyer requirements are translated into a set of configurations that get pushed into the corresponding GLSCs to create individual circuits (steps 12 and 13). Next the connections across endpoints are set up for the duration requested by the buyer in her bid (step 14 and 15). Subsequently, available bandwidth and the number of fiber strand counters are updated (step 16). }
\item The buyer is notified about the decision, along with the connectivity information to access the circuit (step 17). On receipt of this message, end-to-end traffic flow can be initiated by the buyers (step 18). The circuits are continuously monitored by the GLSC to create instant backups in case of failure events.
\item Finally, connection tear down simply causes the established circuit to be revoked between the endpoints. When the lease time of buyers end, this process is triggered automatically.
\squishend

\vspace{-0.35cm}
\section{\gf~Implementation and Evaluation} \label{sec:results} 

In this section, we describe an implementation of \gf, which was developed to provide insights on feasibility and performance.  We also describe results of our evaluation of the implementation in the GENI testbed.

{\bf Implementation.} The \gf~system\footnote{Source code for~\gf~will be made available upon publication.}, along with GGC, GLSC, Fiber Exchange, interfaces for buyers and sellers, and monitoring and measurement subcomponents described in \S\ref{sec:methods} were all implemented in Python. Our implementation includes broad functionality for each \gf~component\footnote{Commercial \gf~deployments will be more scalable and robust, and will reflect details of both business and operational requirements.}.  This enables all aspects of the \gf~event sequence and important aspects of performance to be evaluated.

The GGC is designed to efficiently
serve simultaneous requests from multiple buyers in a multi-threaded fashion and has communication interfaces to different entities including buyers, sellers and Fiber Exchange (via GGC). Resource requests from buyers are sent via the buyer interface as {\tt <Endpoint\_A, Endpoint\_B, \#OfStrandsNeeded, BidAmount, Time, CapacityNeeded, ClientName>} tuples in a {\tt json} format. Next, the physical infrastructure information from the sellers are encoded as topology graphs using the {\tt networkx} library and are sent via the seller interface. Provisions are available in \gf~for both fiber providers and customers to update seller and buyer information respectively. Finally, bid amount and client information extracted from the resource requests are sent to the Fiber Exchange. We note that all the data as well as messages communicated using the aforementioned interfaces are both compressed and encrypted.

Auctions are run at the Fiber Exchange, which implements both GSP- and VCG-based models, and the winner is determined. The winner information from a given auction is communicated to the GGC using the interface specific to Fiber Exchange. The GGC further communicates the winner information to individual GLSC locations.  A GLSC, as noted in \S\ref{sec:methods}, is similar to the GGC albeit with a restricted set of functions and is multi-threaded to improve efficiency. Specifically, it monitors resources using the {\tt ping} tool, transmits resource information to the GGC through an interface to the Fiber Exchange, and uses infrastructure-specific libraries for creating and pushing configurations to physical infrastructure (as explained below). For our experiments, both GGC and GLSC reside on a Macbook Pro laptop equipped with Intel's i5 processor and 4GB RAM.

{\bf Experimental testbed.} We demonstrate and evaluate the \gf~system through deployment in the Global Environment for Network Innovations (GENI) testbed~\cite{Berman20145}. GENI enables relatively controlled testing across a homogeneous infrastructure.  GENI also offers access to network-based devices that are useful for \gf~tests.   We also developed a GLSC that interfaces with Mininet as the underlying network substrate.  We measured the total time taken to bring a circuit into service using each of these systems and while latencies for setting up \gf-internal components were consistent between GENI and Mininet, circuit creation times in Mininet were very small (on the order of microseconds).  Although the GENI-imposed circuit creation latencies are fairly large compared to those obtained in Mininet, we use in this paper the GENI setting as the basis for our evaluation due to the feasibility of experiments and the realism inherent in its wide-area reach. 
Moreover, while some aspects of GENI are idiosyncratic, the availability of configurable devices along end-to-end paths make it an attractive target for our \gf~demonstration.

In our experiments, the resource requests are randomly generated based on the {\em resource pool} information populated by the sellers in the system. Since we use GENI to evaluate~\gf, infrastructure information
from the GENI resource center is used to populate the resource pool and bootstrap our system. Similarly, we use GENI's {\em stitcher} service~\cite{geniStitcher} to create/tear down circuits across GENI endpoints.

Next, in all our experiments, we use a GSP-based auction to elect a winner. If the buyer wins the auction for fiber resource(s), the global controller in GGC issues a new GENI Resource Specification (RSpec)~\cite{geniRspec}---an XML-formatted configuration file used to reserve network resources in GENI---generation request to GLSC and a new RSpec for circuit creation/revocation is created at the GLSC. This configuration file is pushed into the GENI infrastructure {\em only} if the GLSC monitoring a specific set of resources in a geographic location has determined that those resources are continuously available.  In our experiments, we use simple active probes using {\tt ping} to determine availability, and set the monitoring interval to 1s. We note that our approach of monitoring resources is based on ideas borrowed from prior efforts~\cite{ron01, sureroute}.  Furthermore, the monitoring interval is tunable and can be changed by any entity deploying \gf.


\begin{figure*}
  \begin{minipage}[b]{0.31\textwidth}
    \centering
\scalebox{0.7}{
\begin{tabular}{|c|c|c|} \hline
Number of 	& 	Configuration 	&	Circuit \\
Links		&	Generation (s)	&	Provision (s)\\ \hline
 1 			& 	0.124	 	&	19	\\
 2 			& 	0.116 		&	22	\\
 3			& 	0.107		&	21	\\
 4			&	0.148	 	&	25	\\
 5			&	0.126	 	&	24	\\
 10			&	0.112	 	&	33	\\
 20			&	0.119	 	&	35	\\
 30			&	0.120	 	&	37	\\
 40			&	0.112	 	&	47	\\
 50			&	0.121	 	&	54	\\ \hline
\end{tabular}
}
     \captionof{table}{Configuration generation and provision times on scaling the number of links in \gf~system.}
         \vspace{-0.4cm}
             \label{tab:simpleScale} 
    \end{minipage}
    \hfill
    \begin{minipage}[b]{0.68\textwidth}
    \centering
    \epsfig{figure=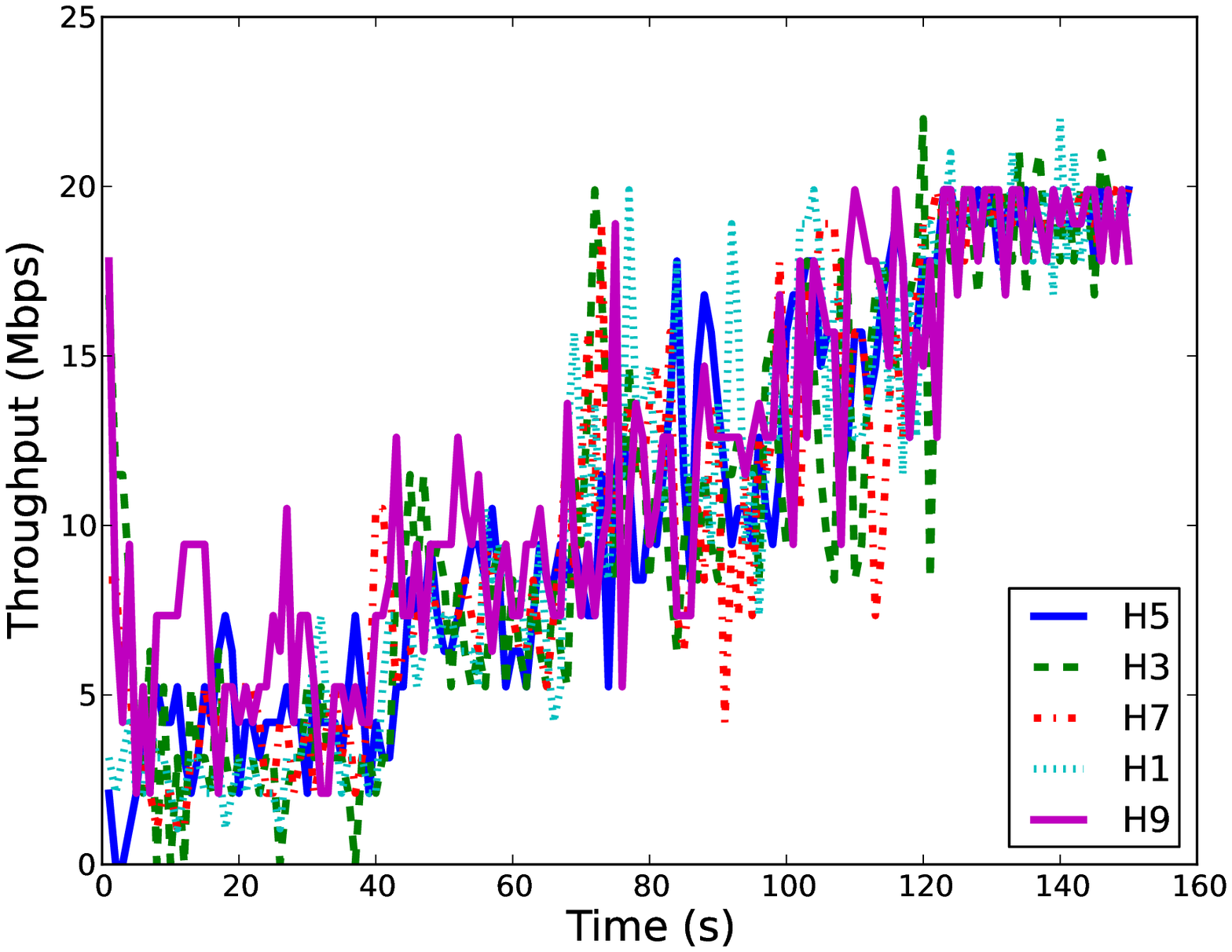,width=6cm}
    \epsfig{figure=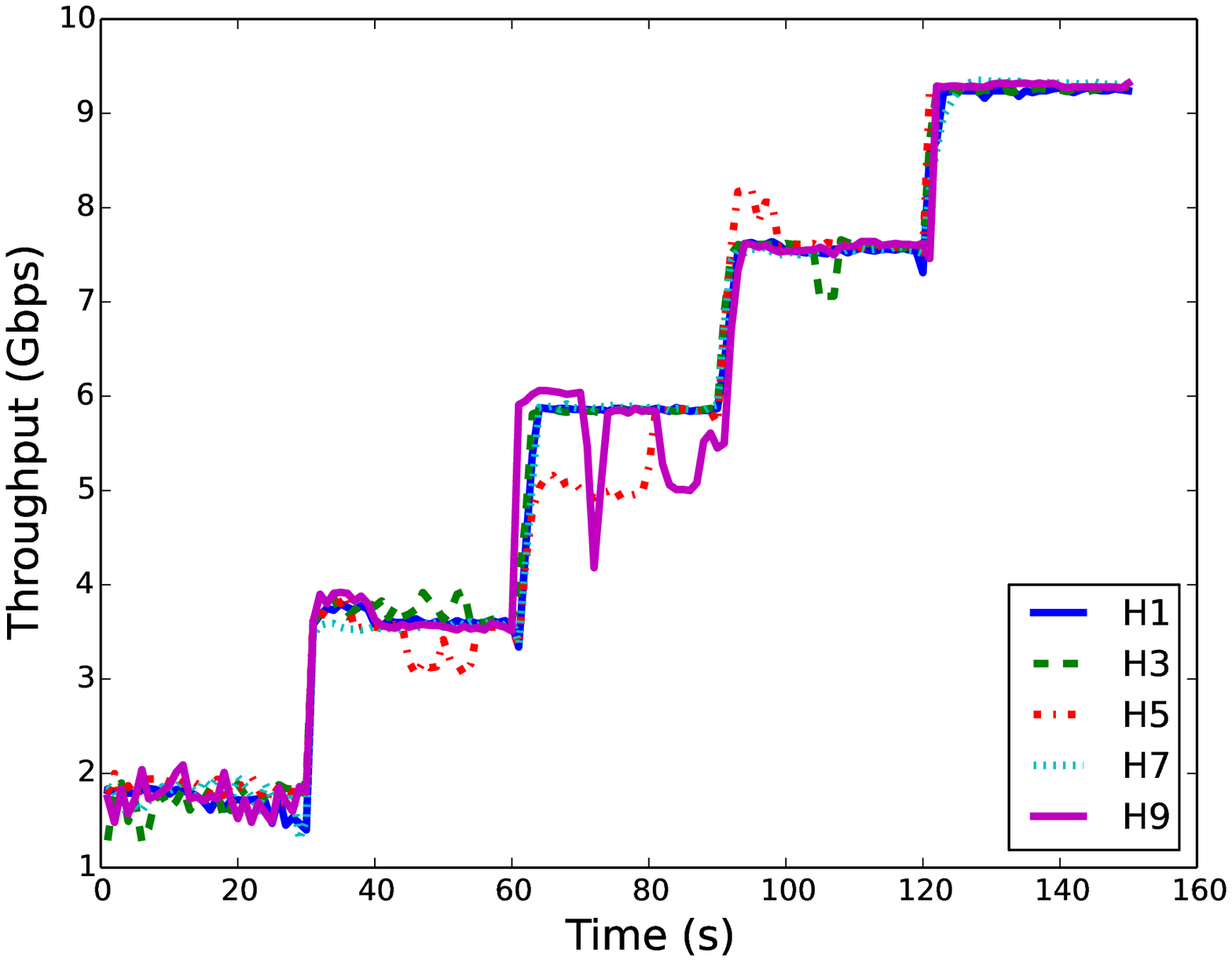,width=6cm}
    \vspace{-0.3cm}
    \captionof{figure}{Performance improvements achieved using~\gf~on GENI (left) and CloudLab (right) testbeds.}
    \vspace{-0.3cm}
    \label{fig:performanceSimple}
  \end{minipage}
  \end{figure*}

The GLSC assigns an available resource in a particular location to satisfy a provisioning request.  If the requested resource is unavailable (as determined using the monitors at GLSC locations) or if the request failed due to unavoidable errors ({\em e.g.}, hardware failure), the next resource at the location is assigned to satisfy the request.

%

{\bf Evaluation methodology.} In our evaluation, we start by focusing on the feasibility and scalability of \gf~system. Next, to demonstrate the ability of \gf~to adapt to network dynamics ({\em e.g.}, failures), we run our experiments in an end-to-end, wide-area setting.   Tests consider both the performance and responsiveness of the system in the presence of background traffic. Specifically, our evaluation is organized around four main questions:

{\bf Q1.} Can \gf~effectively scale if multiple links are required on demand?

{\bf Q2.} What are the performance overheads in the \gf~system?

{\bf Q3.} How performant and responsive is \gf~during network outage(s)?

{\bf Q4.} How does the performance of \gf~for provisioning an alternate path in reaction to a failure compare with rerouting overheads, {\em e.g.}, using OSPF?

\subsection{Scalability of~\gf} To assess the scalability of~\gf, we increase the number of links in a simple dumbbell topology with two nodes.

In this experiment, the two endpoints (or node pairs) are located at two different geographic locations. We repeated the scaling experiments 5 times with different node pairs that are selected randomly from GENI nodes~\cite{geniMap}, at different locations and at different times of the day.

Table~\ref{tab:simpleScale} shows the averages of time taken (in seconds) to generate configuration files and to provision the circuits when increasing the number of links between the dumbbell endpoints for 5 runs of the scaling experiment. The time taken to generate the configuration is about 120ms on average, independent of the number of links.  The time taken to provision circuits from scratch ranges from 19s for one circuit to within a minute (54s) for 50 circuits.  We note that these provisioning times depend on characteristics of the underlying physical infrastructure (in this case, GENI) which are outside the control of the~\gf~system.  For a different infrastructure ({\em e.g.}, controlled through modern optical transport gear), these circuit provisioning times would likely differ significantly.


While \gf~requirements indicate scaling to thousands of circuits, the GENI infrastructure limits our ability to experiment at that scale. Thus, we consider these results as ``proof of concept" and intend to continue to investigate scaling in future work. Our expectation is that future cloud-based or distributed versions of the GGC will satisfy the outlined scalability requirements. Apart from improving scalability, such distributed versions of the GGC would also enable the consideration of regional differences between various sellers, buyers, market economies, and geographic considerations.
For example, the north-eastern region may be dictated by the prevailing business needs of customers requiring low-latency paths for financial transactions. Similarly, the west region may be defined by the need for physical diversity of routes across the Rockies.

\begin{figure*}[!htbp]
  \centerline{\epsfig{figure=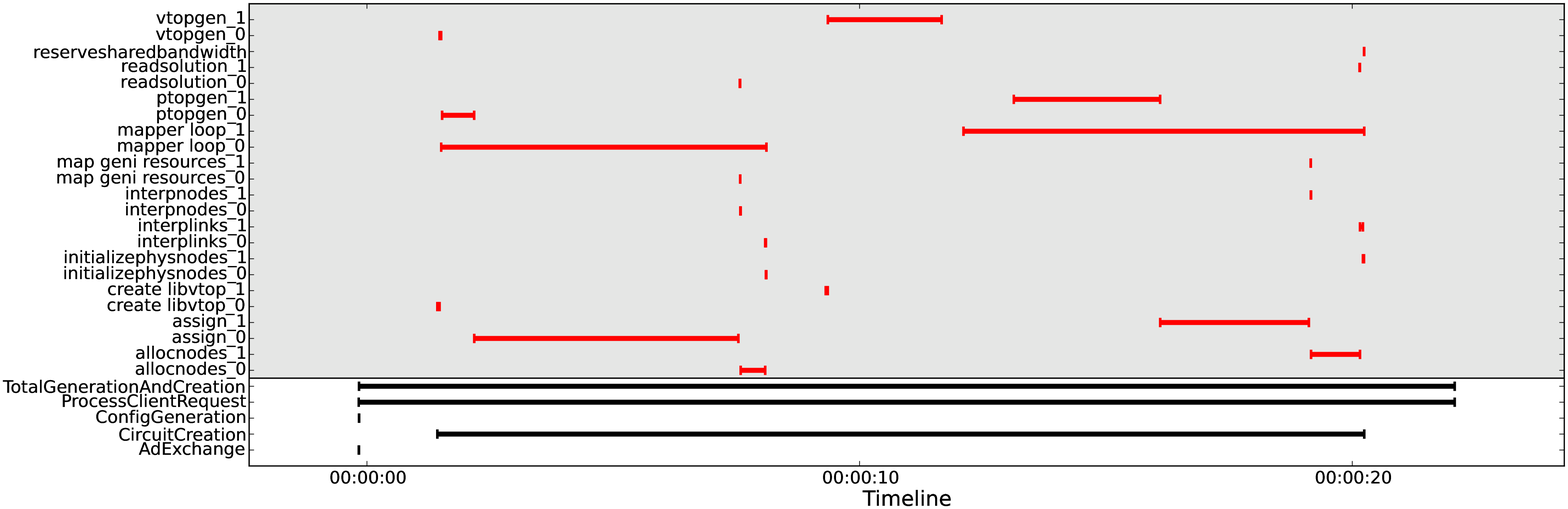,width=16cm, height=6.5cm}}
  \vspace{-0.4cm}
  {\caption{\label{fig:overheadSimple}{Time taken (in seconds) by different components in~\gf~and GENI. Time Timestamps extracted from the {\em spew log} file for dumbbell topology endpoints are marked with "\_0" and "\_1".}}}
  \vspace{-0.4cm}
\end{figure*}

\subsection{Overheads in \gf}

We drill down on the time taken by different components in~\gf~to provision a circuit between two endpoints and quantify the overhead in the~\gf~system versus the underlying network substrate. Specifically, we measure the time spent to generate the configuration files, provision the actual circuits between node pairs, determine the winners of the auction at Fiber Exchange, and total response time to process a buyer's resource request.

Since our measurement framework is opaque to the underlying network gear in GENI, measuring the time taken by individual components ({\em e.g.}, hardware, configuration software, etc.) that are used for circuit provisioning/tear down is beyond the control of \gf. This calls for integration of intuitive measurement methods into our system to effectively measure the \gf~overhead. To that end, our measurement framework utilizes information from GENI {\em spew log} files that are emitted during circuit provisioning to quantify the overheads in the underlying network substrate. Specifically, we extract information such as timestamps and debug messages from the log files to tease out the overheads in GENI versus the overheads in \gf.

Figure~\ref{fig:overheadSimple} depicts the time taken by different modules as reported by our measurement framework, which is available as part of the~\gf~system. Timestamps extracted from spew log files that correspond to GENI infrastructure are shown in red and are marked with a grey background. Processing time taken by individual \gf-specific components including Fiber Exchange (177ms), configuration generation (124ms), circuit creation (18.813s) and client requests (22.245s) for provisioning one circuit
are also shown. Next, we map the circuit creation process into individual GENI-specific functions using the spew log file in the measurement component to account for testbed---in particular, GENI-specific---overheads.
We note that the predominant overhead is caused by {\em mapper loop} function which encompasses other functions including {\em ptopgen}, {\em assign} and {\em interpnodes}, {\em interplinks}, and {\em allocnodes}.
Overall, we observe from Figure~\ref{fig:overheadSimple} that the circuit creation process is responsible for the bulk of the total time required, and that the~\gf~system itself introduces little latency (just over 300ms).  Again, we observe that this inherent latency is completely dependent on the underlying network substrate---an observation consistent with anecdotal evidence from a service provider~\cite{anecdoteAK}.

\subsection{Performance of \gf}

In this experiment, we demonstrate the performance gains---specifically, improvements in throughput---that can be achieved when incrementally adding physical capacity using \gf. For this analysis, we reused the dumbbell topology from earlier experiments, adding an
an {\tt iPerf} server and client at each end point.
Next, we bootstrapped the experiment with five hosts on either side of the bottleneck link, creating five different TCP flows.


To show the performance benefits of \gf, we scale the number of links between the dumbbell topology endpoints by dynamically provisioning a new circuit every 30s. Figure~\ref{fig:performanceSimple}-(left) shows the improvements in performance on scaling the number of links. At the start of the experiment, {\em i.e.}, during the initial 30s, all five flows contended heavily for the bottleneck link
and the average effective throughput, as observed from H1, is $\sim$4Mbps. Upon provisioning two additional links at 30s and 60s, the throughput increases to $\sim$8Mbps and $\sim$12Mbps respectively. On further addition of a link at 90s, an average throughput of $\sim$16Mbps is achieved. Finally, on yet another addition of a link at 120s (leading to a total of 5 links between the dumbbell endpoints), an average effective throughput of $\sim$20Mbps is achieved by all the five competing flows.

\edit{We repeated the experiment (above) on CloudLab~\cite{cloudlab} using the same GENI RSpec, by changing the capacity to 10Gbps. The results are depicted in Figure~\ref{fig:performanceSimple}-(right). Similar to Figure~\ref{fig:performanceSimple}-(left), all five flows contended heavily for the bottleneck link initially and throughput across is $\sim$1.7Gbps. At 30s and 60s two additional links were provisioned, which increased the throughput to $\sim$3.7Gbps and $\sim$5.3Gbps respectively. On further addition of a link at 90s, an average throughput of $\sim$7.6Gbps is achieved. Lastly, an average effective throughput of $\sim$9.55Gbps is achieved by all the five competing flows on provisioning the fifth link at 120s. From this result, we make two key observations: {\em (1)} \gf~scales effectively on links with larger bandwidths {\em without} any performance degradation and {\em (2)} \gf~is generic and adaptable to different networking substrates. These results, apart from showing the efficacy of \gf, also demonstrate the kinds of performance gains that could be achieved using \gf.}

\begin{figure*}[!htb]
\begin{centering}
\epsfig{figure=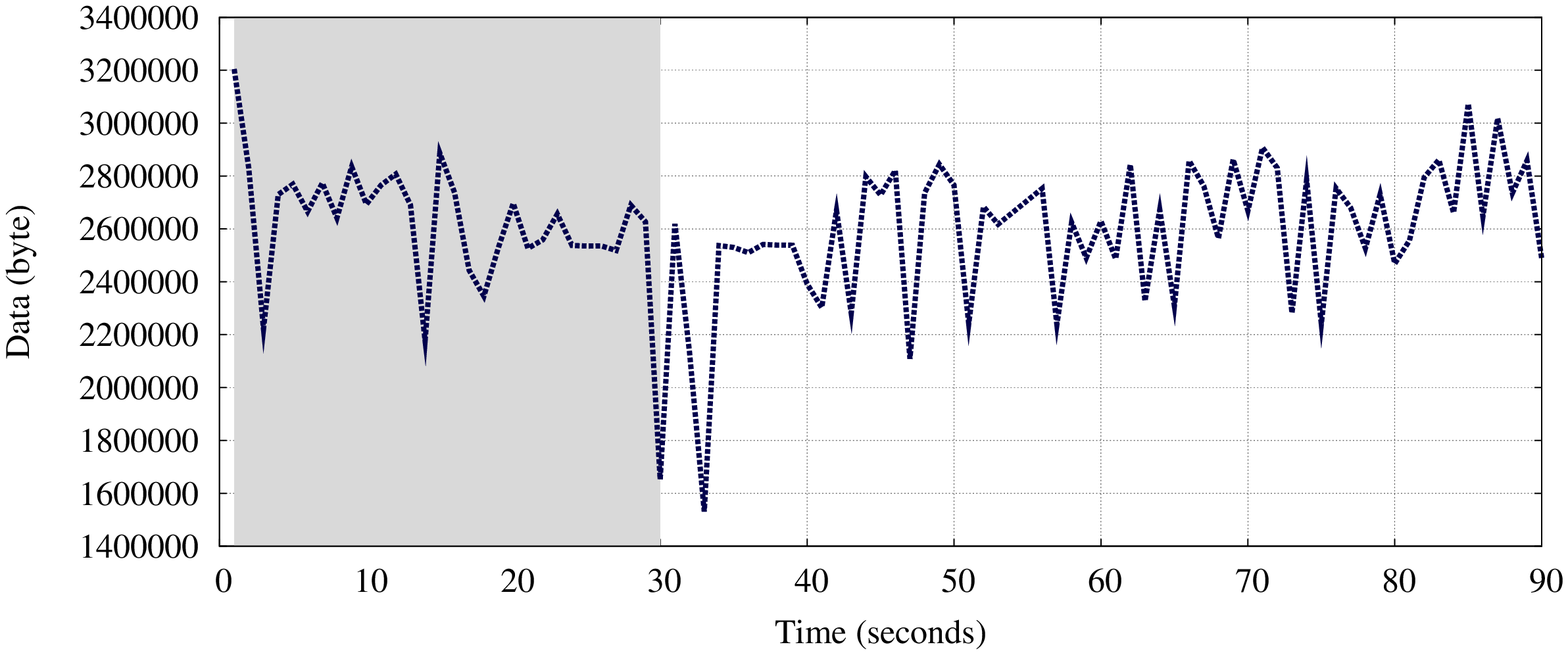,width=7.5cm}
\epsfig{figure=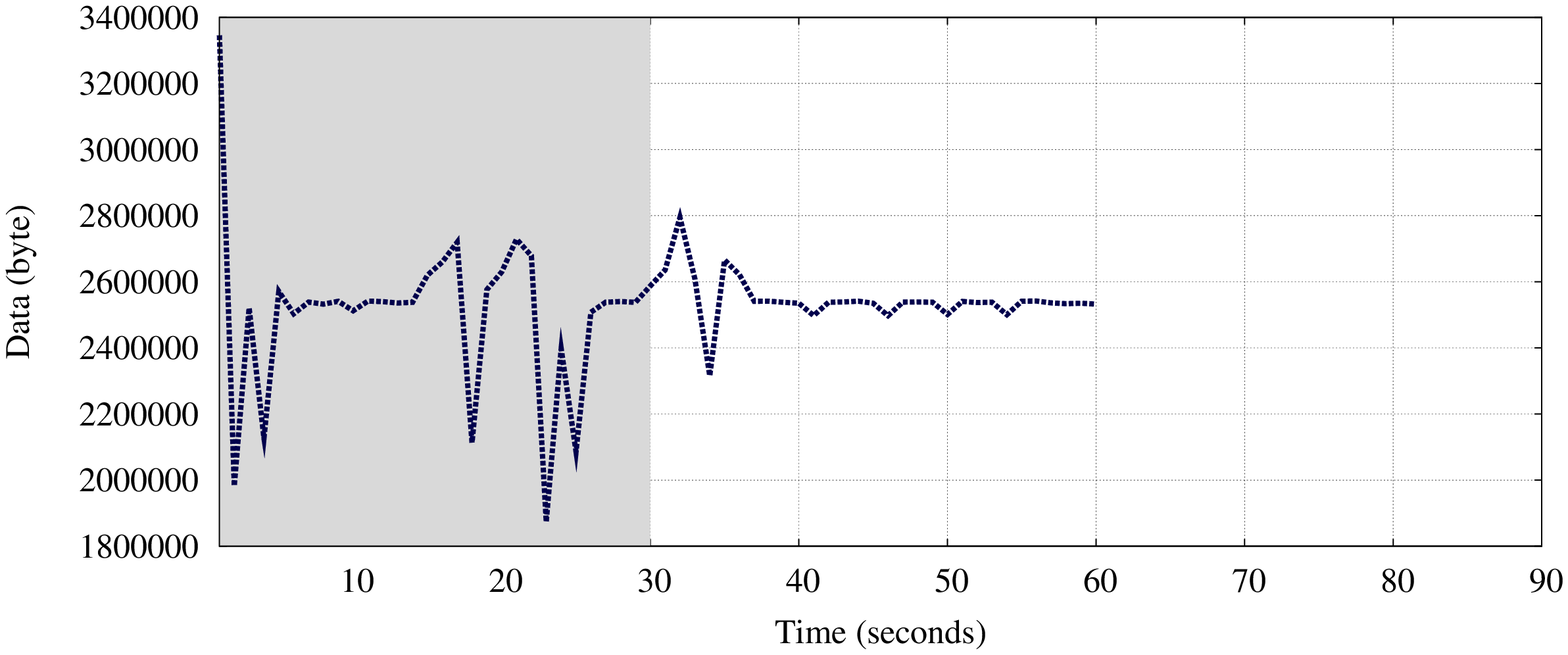,width=7.5cm}\\
\epsfig{figure=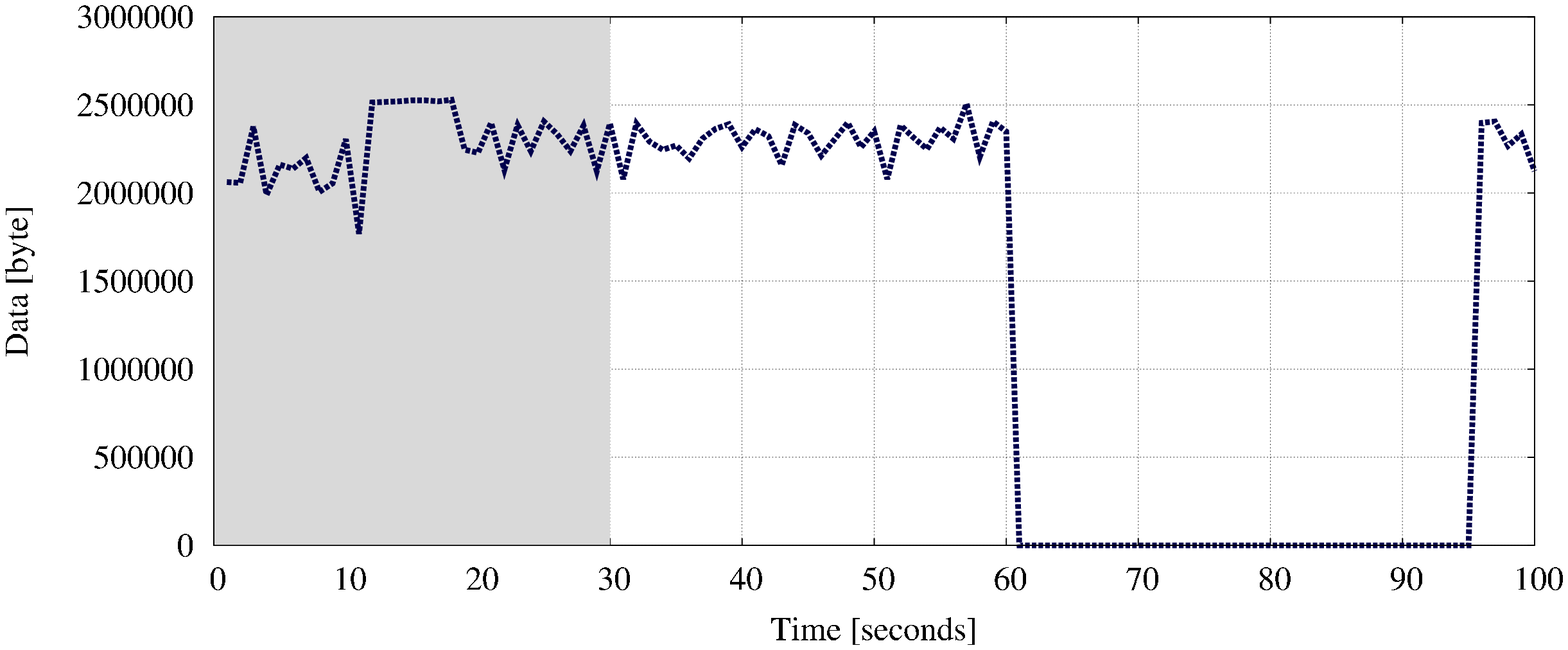,width=7.5cm}
\epsfig{figure=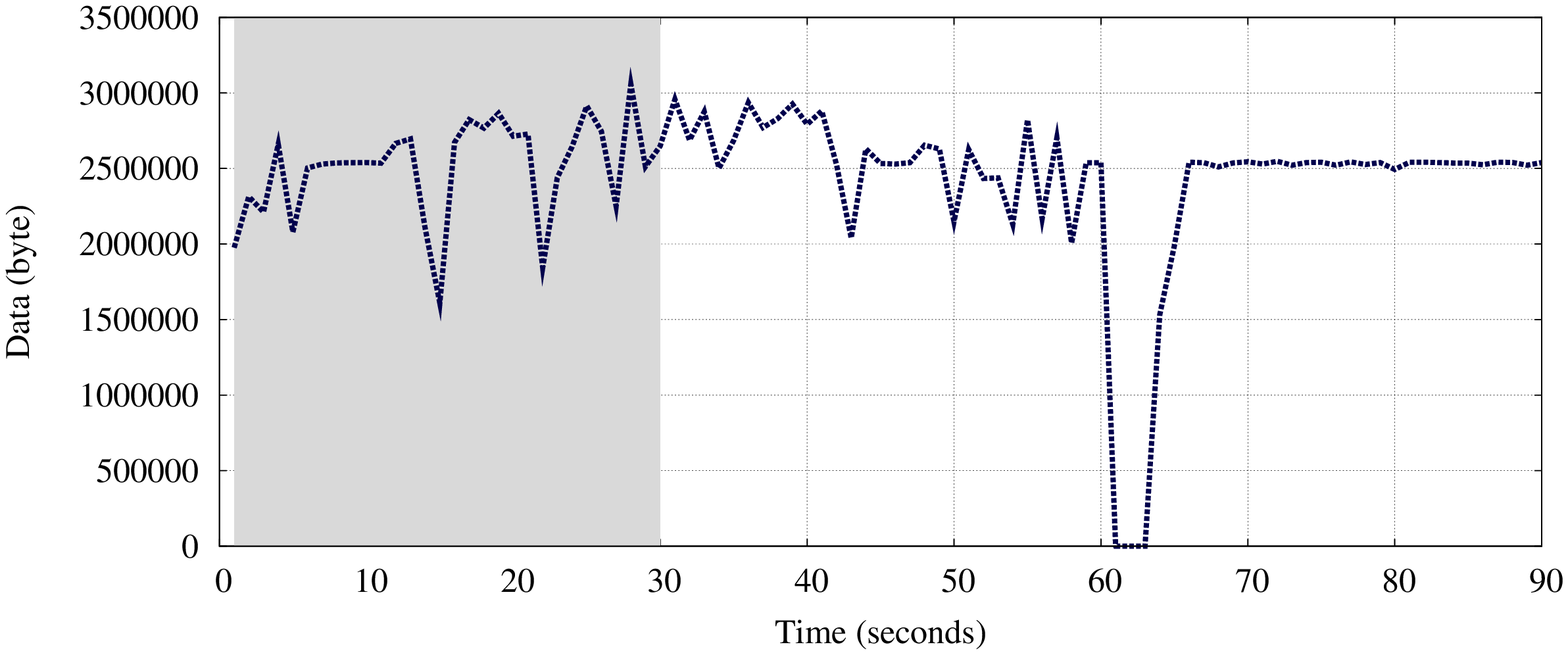,width=7.5cm}

\vspace{-0.2cm}
  {\caption{\label{fig:outageTput}{Throughput (bytes per second) results from dynamic outage detection and recovery experiments.  The warmup phase of each experiment is shown with grey background. Plots shown for no failures (top left), with failures but no backups (top right), with failures and backup using OSPF (bottom left) and \gf~(bottom right).}}}
\vspace{-0.4cm}
\end{centering}
\end{figure*}

\subsection{Effectiveness in the Face of Outages}

Finally, we show how \gf~could be effectively used to provide backup physical connectivity during network maintenance and/or outage events. We start with one link in the dumbbell topology and run an {\tt iPerf} server and client to generate traffic for 90 seconds. The first 30s is the warmup phase to account for TCP artifacts like congestion control.  Next, we manually introduce a {\em link failure event} at the 60th second on the link between the dumbbell endpoints by using the {\tt tc} (traffic control) command and disrupt connectivity in different ways.  For each experiment, we measure and show the throughput (in bytes per second).

{\bf Scenario 1: No failures.} We begin our evaluation by showing the best case scenario where there is no failure event introduced between the dumbbell endpoints.  The top-left plot of Figure~\ref{fig:outageTput}
shows throughput
as observed from the iPerf sender. In this scenario, a total data of 1.88Gb is transferred across the network and the throughput observed is 20.78 Mbps.

{\bf Scenario 2: No backup solution.} Next, we show the effect of a link failure event {\em without} any instantaneous and reactive backup solution in this scenario. This is the worst case scenario. The top-right plot of Figure~\ref{fig:outageTput}
depicts the throughput
for this situation
The connection between the endpoints stalled at the 60th second.  Furthermore, the total data transferred dropped to 1.23Gb, with an average throughput of 20.19Mbps up to the time of failure.

{\bf Scenario 3: Using OSPF-based backup.} Third, we evaluate an OSPF-based backup solution to reroute traffic during the link failure event. In this scenario, {\tt hello} and {\tt dead} intervals are set to 10s and 40s respectively, which are off-the-shelf default values for OSPF.  In this scenario, we used one additional link as a backup to reroute traffic. Also, the experiments were run for 100s to illustrate OSPF's recovery.

The bottom-left plot of Figure~\ref{fig:outageTput}
shows the throughput
using OSPF routing to reroute traffic. We observe a lag of 36s to re-establish connectivity using OSPF-based backup\footnote{When the measurement was taken, the {\tt wait} time interval was 4s. Hence an OSPF-backup was initiated at the ${dead - wait}^{th}$ second.} with a total data transfer of 1.26Gb at 13.04Mbps.  These results are as bad as the no-backup scenario 2.

For the experiment in this scenario, as noted above, we use the default values for time intervals. These values are not proscriptive but are used by service providers in traditional OSPF settings. An alternative way is to reduce the timer hello and dead timer values. However, anecdotal evidence shows that the configurations generated from reduced timer values can be sub-optimal and can result in route flaps~\cite{routeFlap}. In addition, since we use {\em quagga}-based routers
at the network endpoints, to the best of our knowledge, there are no known implementations for mechanism like {\em fast reroute}~\cite{FastRR} and {\em fast hello}~\cite{FastHello}. We intend to evaluate these solutions against \gf-based backup solution as part of future work.

{\bf Scenario 4: Using \gf-based backup.} Finally, in this scenario, we outline the efficacy of a \gf-based backup solution. Specifically, we show how the link failure event introduced at 60th second is rapidly detected by the GLSC, which monitors every network provisioned resource associated to it (by default, every second).  In short, as soon as the failure event is detected, a new link is provisioned by the GLSC thereby initiating a backup.

The bottom-right plot of Figure~\ref{fig:outageTput}
shows the throughput
as the circuit is provisioned using~\gf~on detecting a link outage (at around 60s). During this scenario, a total data of 1.76Gb is transferred across the network at rate of 19.48Mbps.

The GLSC took 1s to detect the link failure event and another 240ms to provision/activate a link in the existing {\em shared vlan}~\cite{sharedVLAN} configured through the GENI infrastructure, and reroute flows via the newly created path. This results in a 28x faster recovery than the OSPF-based scenario. Since, for this experiment, we used a TEQL-based load-sharing technique~\cite{teqlLB} while provisioning circuits between the dumbbell endpoints, links are effectively aggregated and backup creation is rapid.  While the latency of activating the backup link (240ms) is GENI-infrastructure-specific, it is similar to switching times found in published specifications from commercial optical networking gear, {\em e.g.},~\cite{infineraDatasheet}.

While the monitoring interval employed by a GLSC is tunable and the physical infrastructure imposes unavoidable latency in the provisioning process, our results illustrate how \gf~could be used to quickly recover from network outages with minimal impact on user traffic. For example, a video streaming application with modest buffering would not perceive any glitch, and for chat, interactive shell, and other realtime applications, the impact would be short-lived.  Lastly, for web traffic, the waiting time to lose a user has been observed to be $\sim$4s~\cite{surveyNNGroup}. Even with a more stringent two-second rule for webpage load times~\cite{twoSecondRule}, the \gf~system can sufficiently provision a backup path.

\section{Related Work} \label{sec:relatedwork} 


{\bf Infrastructure provisioning.} 
In the context of datacenter and WAN settings, infrastructure provisioning has been of interest to both industrial and academic communities~\cite{jin2016, laoutaris2011inter, jain2013b4, hong2013achieving}. \editJocn{SDN-based provisioning approaches include B4~\cite{jain2013b4}, SWAN~\cite{hong2013achieving}, Owan~\cite{jin2016}, and others~\cite{singh2017run, channegowda2013software, giorgetti2015dynamic, kilper2017optical, xiong2018sdn}, each of which aims at improving the utilization of inter-datacenter and wide area networks. A survey of related efforts are available here~\cite{thyagaturu2016software, alvizu2017comprehensive}.} We posit that deployment of such efforts along with acquiring access to physical paths (via IRU or \gf) between DCs has the potential to produce better performance results than considering either of these solutions in isolation. In particular, we argue that such an environment, which considers provisioning and access to physical paths in tandem, can facilitate improvements at the physical layer~\cite{xu2013provisioning},
network layer optimizations~\cite{kumar2015bwe}, and cross-layer enhancements, {\em e.g.},~\cite{chiu2012architectures,guan2007cost}.


{\bf Internet economics.} Incorporating pricing models for networks has been of interest to researchers since the Internet's infancy~\cite{shenker1996pricing, kelly1998rate, ma2010internet}.
Recently, many efforts have focused on increasing revenues for service providers and customer satisfaction via flexible pricing models. For instance, Jalaparthi {\em et al.}~\cite{jalaparti2016dynamic} accommodates both deadlines and demands into a time-dependent pricing model to create Pretium, a framework which considers economics and traffic engineering issues in tandem. Similarly, a pricing model for transit ISPs based on tiers and traffic demand is proposed in~\cite{valancius2011many}.

The auction model in \gf~is motivated by online auction research in the theory literature. Specifically, we use the classical results on Generalized Second Price (GSP)~\cite{edelman2007internet} or Vickrey-Clarke-Groves (VCG)~\cite{Vickrey1961,Clarke1971,Theodore1973} in our framework.
Furthermore, several industrial efforts on infrastructure economics include bandwidth markets ({\em e.g.}, Enron~\cite{enron}), spot pricing markets ({\em e.g.}, Invisible Hand Networks~\cite{invisibleHand}), and fiber arbiters ({\em e.g.}, IXReach~\cite{ixreach}).
In particular, IXReach (which was acquired in 2015 by IIX, Inc.~\cite{ixreachAcquisition} which in turn was renamed as Console~\cite{consoleRename}) provides the ability to expand network footprint at locations that are of interest to service providers $\grave{a}~la$ GreyFiber.

\vspace{-0.35cm}
\section{Summary and Future Work} \label{sec:summary} 

Our work is motivated by the fact that market forces and technology trends have evolved to the point where alternatives to the decades-old methods for gaining access to physical network infrastructure (dark vs. lit fiber) are now feasible.  In this paper, we describe \gf, which is designed to enable wide area connectivity as a service, similar to the way that cloud computing has enabled computation-based service offerings that have had a transformative impact.  The objective of \gf~is to offer flexible access to fiber-optic paths between \editJocn{end points ({\em e.g.,} datacenters and/or colocation facilities)} over a range of timescales, and through a Fiber Exchange, which makes this connectivity available to the highest bidders.
We design and deploy an instance of a \gf~system and evaluate it. We show that circuit provisioning time scales roughly linearly with the number of links, that overheads are tightly coupled with the infrastructure under \gf~control, and that \gf~could be effectively used to improve path performance and recover from outages.
While 
our results demonstrate the efficacy of our \gf~design, there is much to be done in future work to develop the core concepts into reliable, high performance systems that deliver wide area connectivity as a service.  In on-going work we are developing partnerships with service and equipment providers toward the goal of deploying \gf~in a live environment.  One of the key aspects of this work is to push functionality as close to the physical layer as possible in order to reduce provisioning latency. At the same time, we plan to address scaling and distributing the GGC. We also plan to expand our cost, pricing and deployment analyses in order to assess the feasibility of wide area connectivity as a service in a range of markets. 

\ifCLASSOPTIONcaptionsoff
  \newpage
\fi

\vspace{-0.25cm}
\balance
\bibliographystyle{IEEEtran}
\bibliography{paper,paperP}








\end{document}